# Evolution of the electronic and superconducting properties of Re-based quinary high-entropy alloys under chemical and physical pressure

Thiagarajan Maran [1], Yoshiya Uwatoko [2, 3], Sathea Suweatha M N [1], Sonika Jangid [4], R.P. Singh [4], Arumugam Sonachalam [1,†], and Sathiskumar Mariappan [1, 2, 5,*]

[1] Centre for High Pressure Research, School of Physics, Bharathidasan University, Tiruchirappalli, 620024, India.
[2] Institute for Solid State Physics, The University of Tokyo, 5-1-5 Kashiwanoha, Chiba 277-8581, Japan.
[3] Faculty of Science and Engineering, Tokyo City University, Setagaya, Tokyo 158-8557, Japan.
[4] Indian Institute of Science Education and Research Bhopal, Bhopal 462066, India.
[5] Department of Physics and Astronomy, Seoul National University, Seoul 08826, Korea

E-mail: † sarumugam1963@yahoo.com, *sathiskumarm66@gmail.com

**Abstract:**

We report a comparative study of chemical- and physical-pressure effects on the electronic and superconducting properties of the Re-based quinary high-entropy alloys (HEAs) $[Nb_{0.67-x}Re_x][TiZrHf]_{0.33}$ (x = 0.10, 0.20, and 0.56). Increasing Re concentration suppresses the superconducting transition temperature $T_c$ from 5.4 to 3.9 K while producing positive, composition-dependent cocktail-effect ratios. Field-dependent transport and magnetization establish all three compositions as strongly type-II superconductors and identify x = 0.20 as the most distinctive composition, with the largest upper critical field and Ginzburg-Landau parameter and a Maki parameter close to unity. Most importantly, the pressure coefficient of $T_c$ changes sign across the bcc-hcp structural change: $dT_c/dP$ is positive for the bcc x = 0.10 and 0.20 samples, with values of approximately 0.018 and 0.053 K/GPa, respectively, but negative for the hcp x = 0.56 sample, with a value of approximately -0.033 K/GPa. This systematic contrast within a single chemically related alloy series establishes a robust structure-associated superconducting response and identifies crystal structure as a key organizing variable under compression. Metallic transport and superconductivity remain robust up to approximately 8 GPa in all three compositions. These results reveal that chemical substitution and hydrostatic compression are complementary but nonequivalent routes for tuning superconductivity in Re-based HEAs.

**Introduction:**

Conventional alloys exhibit useful properties, including good thermal & electrical conductivity, high tensile strength, shape memory effects, low weight-to-strength ratio, malleability, ductility, and corrosion resistance [1–7]. They also host phenomena such as mictomagnetism, tunable quantum stability, and superconductivity [8,9]. These alloys are generally composed of two or three elements, with one principal element occupying the largest atomic fraction. This paradigm has been altered over the past two decades by the discovery of high-entropy alloys (HEAs). Generally, HEAs focus on mechanical and industrial applications due to their high fracture toughness at low temperatures, higher strength arising from low diffusion of the constituent elements, and higher corrosion resistance, among other properties [10].

Unlike traditional alloys, HEAs are typically composed of five or more principal elements, each with a minimum concentration of 5 at.% [11] with high configurational entropy ($S_{config}$) due to the random atomic arrangement on the crystallographic positions [10,12] . Notably, in HEAs, entropy plays a crucial role in stabilizing disordered solid-state solution (SSS) phases, such as simple crystal structures like Body-Centered Cubic (bcc), Face-Centered Cubic (fcc), and Hexagonal Close-Packed (hcp) [13]. In addition, HEAs have attracted attention because of properties such as irradiation resistance (NiCoFeCrMn) [14], soft magnetism (CoFeMnNiAl) [15], potentiality in thermoelectric (PbSnTeSe) [16], and catalytic properties (CoMoFeNiCu) [17]. Surprisingly, some HEAs exhibit superconductivity (SC), even with highly disordered structures.

Following the first report of a quinary high entropy alloy (HEA) superconductor, $Ta_{34}Nb_{33}Hf_{8}Zr_{14}Ti_{11}$, with a superconducting transition temperature ($T_c$) of 7.3 K in 2014 [18], several HEA superconductors with bcc and hcp crystal structures have been published, with a heavily skewed ratio of reports. The dependency of superconductivity on structural polymorphism and the interplay between chemical disorder and structure has also been reported in the $(V_{0.5}Nb_{0.5})_{3-x}Mo_{x}Al_{0.5}Ga_{0.5}$ sample [19]. Remarkably, for x=0.2, the observed $T_c$ of 10.2 K and upper critical field, $\mu_0 H_{c2}(0)$, of 20.1 T are the highest reported values among all HEAs at ambient pressure. Similarly, Re-Os-W-Mo-Ta [20] and Hf-Ta-Nb-Mo-W [21] demonstrate the influence of chemical pressure on superconducting characteristics, including strong-coupling superconductivity ($\Delta C_{el}/\gamma T_c$ > 1.43) and fully gapped *s*-wave pairing. For bcc-structured HEAs, $Ta_{1/6}Nb_{2/6}Hf_{1/6}Zr_{1/6}Ti_{1/6}$ has been reported to exhibit strong correlated and strongly coupled s-wave superconductivity [22]. These reports show that superconductivity can be tuned through structure, VEC, and chemical pressure. Compared with extensive research on the influence of chemical pressure, investigations into external pressure effects remain notably limited, particularly in superconducting HEAs.

Physical pressure (*P*) is a well-established tool for investigating the evolution of structural and electronic properties without permanently altering the material's chemical composition. In addition, several studies on HEAs show the vital role of pressure. For example, in $Mo_xCrFeCoNi$, the dominant role of pressure in the phase transition from fcc to hcp, as a result of pressure on the electronic structure, is speculated [23]. Similarly, CrMnFeCoNi exhibits a sluggish pressure-induced transition from fcc to hcp associated with band broadening [24], and a pressure-induced magnetic phase transition has also been predicted [25]. As for the investigation of HEA superconductors under pressure, both bulk [26] and thin film (~600 nm) [27] of TaNbHfZrTi show $T_c$ of ~7.7 K (+ $dT_c/dP$ up to ~ 80 GPa) and ~5.3 K (-$dT_c/dP$ up to ~3.3 GPa), respectively, exhibiting the significance of the bulk sample. Further, robustness of the superconducting phase along with structural integrity, in TaNbHfZrTi up to ~190 GPa [26] and in $(ScZrNbTa)_{0.6}(RhPd)_{0.3}$ up to ~90 GPa [10] was observed with saturation in $dT_c/dP$, which can be correlated with the universal trend of ~30% volume compression. But this universal trend has been defied in quaternary medium-entropy alloy TaNbHfZr, exhibiting a dome-shaped trend with the highest $T_c$ of 15.3 K at ~72 GPa [28]. Remarkably, an irreversible phase transition from hcp to bcc at 44 GPa has been observed in $Re_{0.6}(NbTiZrHf)_{0.4}$ (VEC – 5.9), a sister compound of our studied samples, and reported [29]. Furthermore, Sun and Cava classify HEA superconductors into four categories (Type A to D) depending on factors such as constituent elements' placement in the periodic table, crystallization phase, range of transition temperature ($T_c$), and upper or lower critical field ($\mu_0H_{c1}$ or $\mu_0H_{c2}$) [10].

Motivated by these results, we performed a comparative study of $[Nb_{0.67-x}Re_x][TiZrHf]_{0.33}$ (x= 0.10, 0.20, 0.56) possessing bcc for x = 0.10, 0.20 (Type A) and hcp structure for x = 0.56 (Type D) under ambient and applied pressure conditions. From here on, sample concentrations x = 0.10, 0.20, and 0.56 are denoted as RNZHT 10, RNZHT 20, and RNZHT 56, respectively. To the best of our knowledge, this is the first report on the superconducting behavior of an hcp HEA probed under *P*. Among these three compositions, the ambient-pressure properties of RNZHT 10 [30] and RNZHT 56 [31] have already been reported. Our results show that all three samples undergo a superconducting transition, and RNZHT 10 exhibits the highest $T_c$ of 5.423 K. Moreover, the effect of chemical pressure on the transport and magnetic properties has been studied to probe superconducting phenomena in $[Nb_{0.67-x}Re_x][TiZrHf]_{0.33}$. The high-pressure resistivity measurements up to ~8 GPa, along with their detailed results for all three samples in assessing the influence of hydrostatic pressure on the superconducting phase, are also reported in this article.

**Experimental Details:**

Polycrystalline $[Nb_{0.67-x}Re_x][TiZrHf]_{0.33}$ samples with x = 0.10, 0.20, and 0.56 were synthesized by arc-melting, as reported elsewhere [31]. The electrical resistivity as a function of temperature, $\rho(T)$, was measured at both ambient and high-pressure (HP) using a Quantum Design Physical Property Measurement System (PPMS) down to 1.8 K and in various magnetic fields up to 9 T. The linear four-probe method was used to measure the electrical resistivity. The electrical contacts were made using a high-quality silver paste and Au wire (ϕ 20 μm). A cubic anvil device was used to measure $\rho(T)$ up to a maximum pressure ($P_{Max}$) ~8 GPa. A 1:1 mixture of Fluorinert FC-70 and FC-77 was used as the pressure-transmitting medium to maintain hydrostatic pressure conditions. The Teflon capsule was placed in a pyrophyllite gasket between the six WC anvils. The sufficiently hydrostatic pressure conditions provided by this cubic-anvil configuration, even after solidification of the pressure medium, have been established previously [32,33]. The pressure cell was calibrated using the fixed pressure points from the bismuth calibration [34]. A Magnetic Properties Measurement System (MPMS-XL, Quantum Design, USA) was used to measure the temperature dependence of magnetization, $M(T)$, in both zero-field cooling (ZFC) and field-cooling (FC) modes and field-dependent magnetization, $M(\mu_0H)$, at ambient pressure.

**Results and discussion:**

**1. Transport and magnetic properties under chemical pressure:**

**(a) Electrical transport measurements:**

Figure 1(a) shows the low-temperature region of temperature-dependent resistivity, $\rho(T)$, at 0 T, where a sudden drop in resistivity to zero represents the superconducting transition of $[Nb_{0.67-x}Re_x][TiZrHf]_{0.33}$ (x=0.10, 0.20, 0.56). The onset ($T_c^{on}$), offset ($T_c^{off}$), and mid transition ($T_c^{mid}$) temperatures are measured at the intersection between the normal state resistivity-transition regime-zero resistivity, and 50% of the onset resistivity, respectively. The $T_c^{on}$, $T_c^{off}$, and $T_c^{mid}$ are extracted as shown in Fig.1 (a) and plotted in the inset of Fig.1(a). The values of $T_c$ are in good agreement with the previous reports [30,31]. The decrement in $T_c$ with increasing Re concentration follows the valence electron count (VEC) dependent $T_c$ phase diagram of the 4d metals [35,36]. Although RNZHT 10 and 56 show a sharp transition, a broader superconducting transition width $\Delta T_c = T_c^{on} - T_c^{off}$ of RNZHT 20 indicates enhanced compositional inhomogeneity, disorder, or granularity. This could be due to the fact that RNZHT 20 is the marginal composition beyond which the presence of mixed phase up to RNZHT 33 can be observed [31]. The steadily decreasing trend of $T_c^{mid}$ with Re concentration, as shown in the inset of Fig.1(a), suggests the accompaniment of $T_c$ with structural transition from bcc to hcp with respect to Re fraction may be ascribed to VEC. Taken together, the monotonic suppression of $T_c$, the evolution of VEC, the increase in residual scattering, and the reported structural evolution support a coupled electronic-structural origin of the chemical-pressure response. These results are similar to $(V_{0.5}Nb_{0.5})_{3-x}Mo_xAl_{0.5}Ga_{0.5}$, where $T_c$ attenuates with increasing at. % of Mo associating a structural transition from pure bcc to a mixed phase of bcc and cubic A15 type structure [19].

Figure 1(b) shows the temperature dependence of resistivity, $\rho(T)$, at zero magnetic field of the HEAs RNZHT 10, 20, and 56. The normal-state $\rho(T)$ (above $T_c$) of all samples shows a weak metallic behavior from 300 K to $T_c$. The metallicity of RNZHT 10 and 56 is relatively higher than that of RNZHT 20. However, the resistivity of RNZHT 56 is fivefold that of RNZHT 10, which reflects the combined effects of disorder, electronic structure, and the change in crystal structure. The highest and lowest $T_c$ values, 5.423 K and 3.907 K, were observed at RNZHT 10 and 56, respectively. The decrease in $T_c$ follows the composition-dependent VEC trend and, together with the concurrent structural evolution, supports a coupled electronic-structural suppression of superconductivity [37,38]. The normal-state $\rho(T)$ data were analyzed using the Fermi-liquid relation ($\rho = \rho_0 + AT^2$) in the temperature range 6 K ≤ T ≤ 75 K and fitted as shown in the inset of Fig.1(b). The extracted residual resistivity ($\rho_0$) and electron-electron scattering factor ($A$) as a function of Re concentration are plotted in Fig.1(d). The increase in $\rho_0$ with chemical pressure (Re concentration) suggests an increase in disorder with Re, and the variation of $A$ may reflect changes in low-temperature scattering near the composition where structural evolution has been reported [31]. The relatively small fitted $A$ values indicate a weak $T^2$ contribution to the resistivity; however, $A$ alone does not provide a quantitative measure of electronic-correlation strength. To understand this further, the evolution of the residual resistivity ratio (RRR) calculated using ($\frac{\rho^{300\ K}}{\rho^{10\ K}}$) and room-temperature resistivity are plotted in Fig.1(c). The minimum RRR, broadened superconducting transition, and enhanced upper critical field near RNZHT 20 consistently point to enhanced scattering in the vicinity of the reported structural crossover.

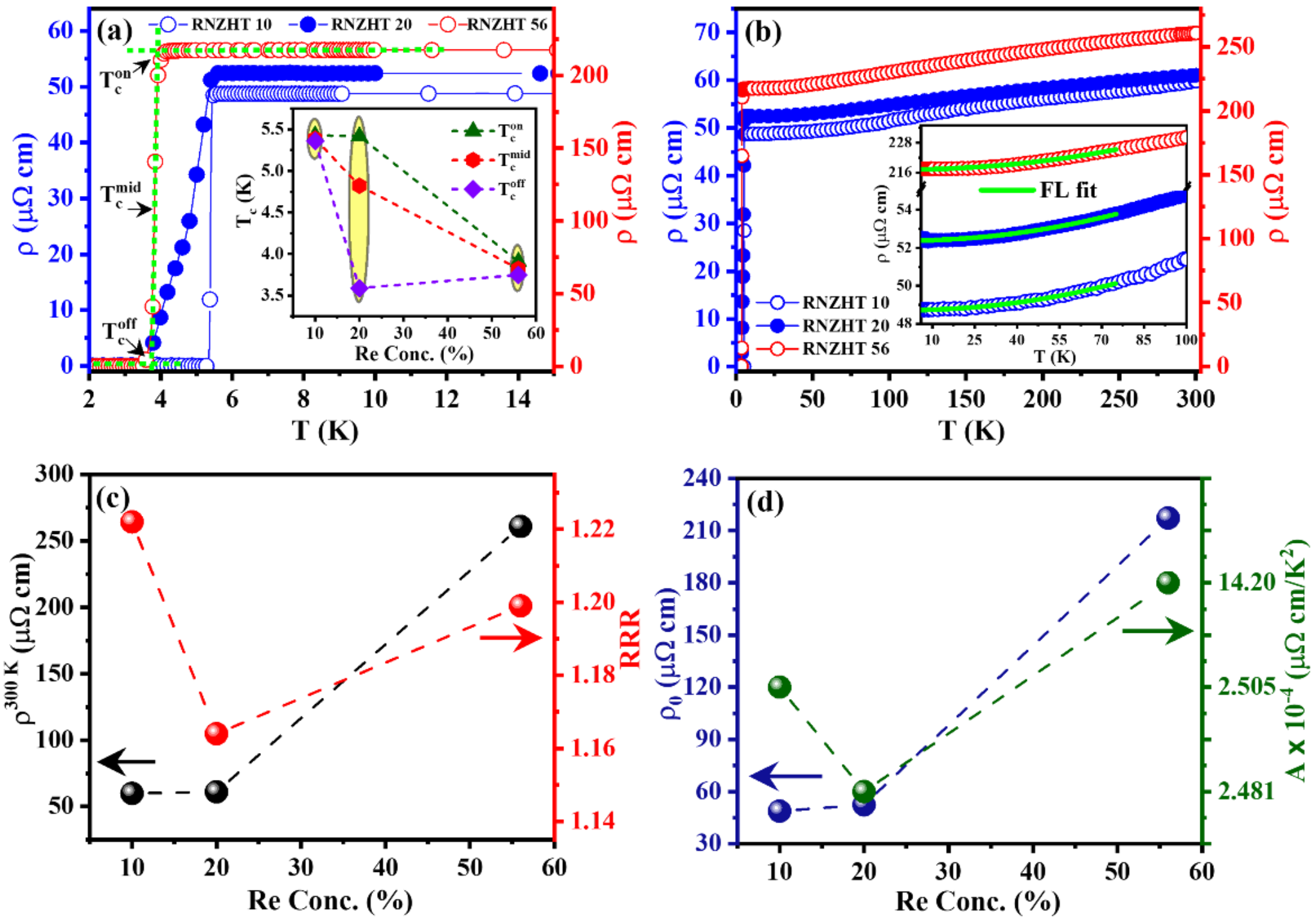

***Figure 1:*** ***(a)*** *shows the magnified $T_c$ region of resistivity vs temperature at zero field for RNZHT 10, 20, & 56 where blue lines corresponds to left axis and red to the right axis* ***(b)*** *shows the resistivity vs temperature for RNZHT 10, 20, & 56 with Fermi Liquid fit (FL fit) for the range of $T_c < T < 75$ K as inset,* ***(c)*** *Resistivity at 300 K (left axis) & RRR (right axis) vs Re concentration %,* ***(d)*** *Residual resistivity $\rho_0$ (left axis) and Scattering factor A (right axis) vs Re concentration % extracted using FL fit.*

### (b) Electrical transport measurement under a magnetic field:

To understand the effect of an external magnetic field ($\mu_0 H$) on superconductivity, the $\rho(T)$ below 10 K was measured under various fixed magnetic fields up to 5 T for all compositions and plotted in Fig.2(a-c). The typical broadening of $T_c$ under various fields suggests the dissipation of vortices. Besides, even though RNZHT 20 has a large $\Delta T_c$, it exhibits a higher upper critical field, $\mu_0 H_{c2}$, than RNZHT 10 and 56. To ascertain the upper critical field at 0 K, $\mu_0 H_{c2}(0)$, and other superconducting derived parameters, we have carefully analyzed the data. We have extracted the $T_c$ for each field and plotted the normalized transition temperature ($T/T_0$) against the $\mu_0 H_{c2}$ ($t$), where $t = T/T_0$, as shown in Fig.2(d). To calculate the $\mu_0 H_{c2}(0)$, the Ginzburg-Landau (GL) relation, $\mu_0 H_{c2}^{GL}(t) = \mu_0 H_{c2}^{GL}(0)[(1-t^2)/(1+t^2)]$, has been fitted (bright green solid lines) on the above-mentioned plots and depicted in Fig.2(d). This clearly suggests the RNZHT 20 and RNZHT 56 have the highest and lowest $\mu_0H_{c2}(0)$, respectively.

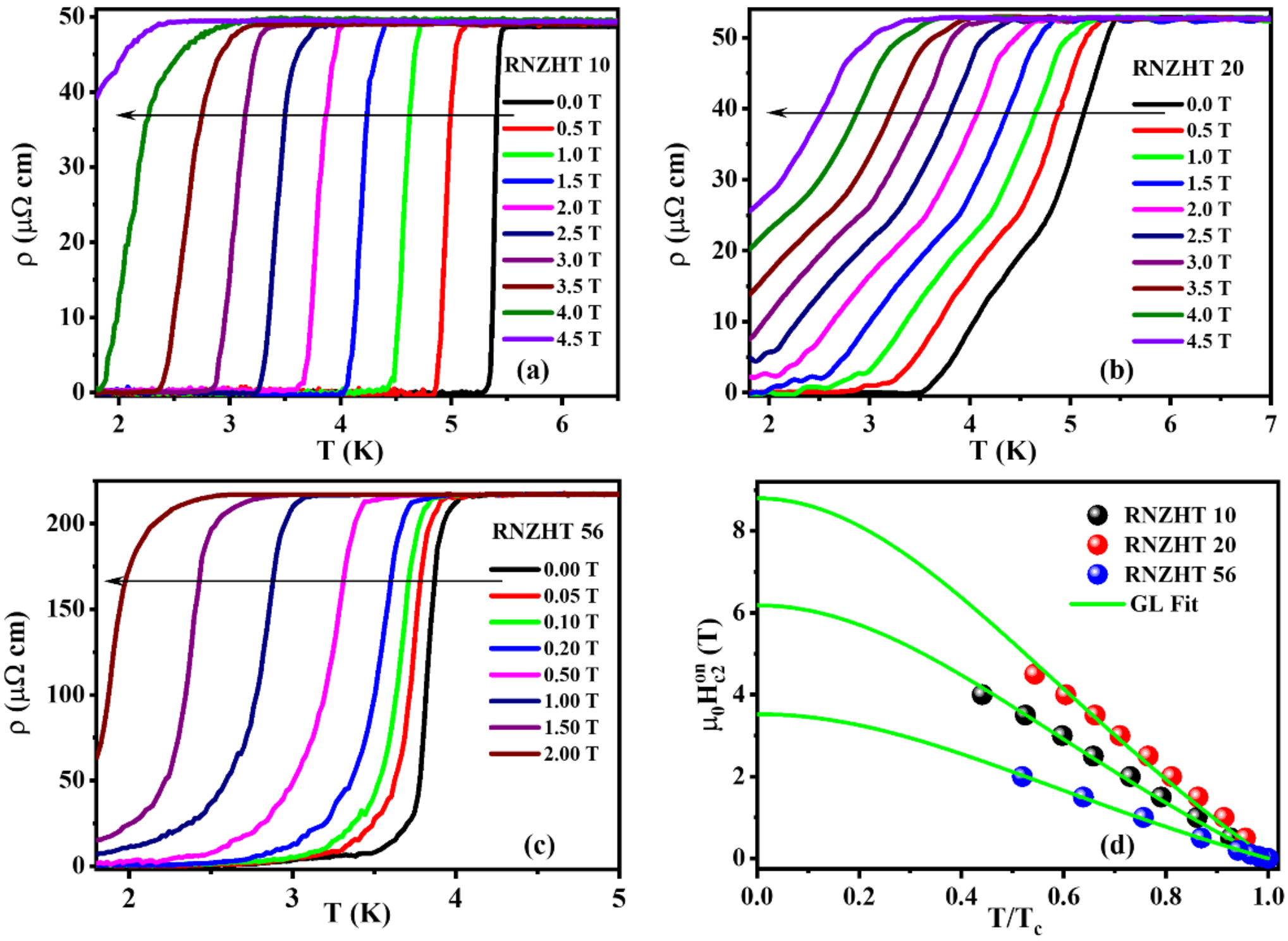

***Figure 2: (a)-(c)*** *show the magnified $T_c$ region of resistivity vs temperature under various fixed magnetic fields up to 4.5 T for all three samples,* ***(d)*** *shows the critical magnetic field $\mu_0H_{c2}$ vs normalized temperature ($T/T_c$) (solid green line represents the GL fit over the plots).*

In addition to $\mu_0H_{c2}^{GL}$, the orbital pair-breaking limit $\mu_0H_{c2}^{Orb}$ has been determined using the single-band Werthamer-Helfand-Hohenberg (WHH) relation, $\mu_0H_{c2}^{WHH}(0) = -0.693\ T_c[d(\mu_0H_{c2})/dT]_{T=T_c}$, where $[d(\mu_0H_{c2})/dT]_{T=T_c}$ is the slope of $\mu_0H_{c2}$ curve at $T = T_c$ shown in Fig.3 (b) and 0.693 is the dirty limit [39]. To understand the magnetic field effect of superconductivity, the Pauli paramagnetic limit has been calculated using the relation, $\mu_0H_{c2}^{P}(0) = 1.86 \times T_c$ [40]. The values of $\mu_0H_{c2}^{GL}(0)$, $\mu_0H_{c2}^{WHH}(0)$, & $\mu_0H_{c2}^{P}(0)$ for all compositions plotted with Re concentration (chemical pressure) are shown in Fig.3(a). The relationship $\mu_0H_{c2}^{\mathrm{P}}(0) > \mu_0H_{c2}^{\mathrm{GL}}(0) > \mu_0H_{c2}^{WHH}(0)$ is consistent with conventional orbital-limited behavior within the single-band WHH framework, but it does not by itself determine the superconducting gap symmetry. Also, no pronounced upward curvatures were observed in $\mu_0H_{c2}(T)$, contrary to the findings in [41,42]. However, this observation alone does not exclude the possibility of multigap superconductivity. Among these three samples, RNZHT 20 shows the highest upper critical field, suggesting the enhancement of $\mu_0H_{c2}(0)$ with chemical pressure (Re concentration), but the structural transition from cubic to mixed phase and further to hcp after RNZHT 20 (x = 0.20) [31] is accompanied by a decrease in $\mu_0 H_{c2}(0)$. Together with the minimum RRR and the broad superconducting transition, the maximum in $\mu_0 H_{c2}(0)$ identifies RNZHT 20 as the composition in which disorder and proximity to the reported structural crossover most strongly influence the superconducting state.

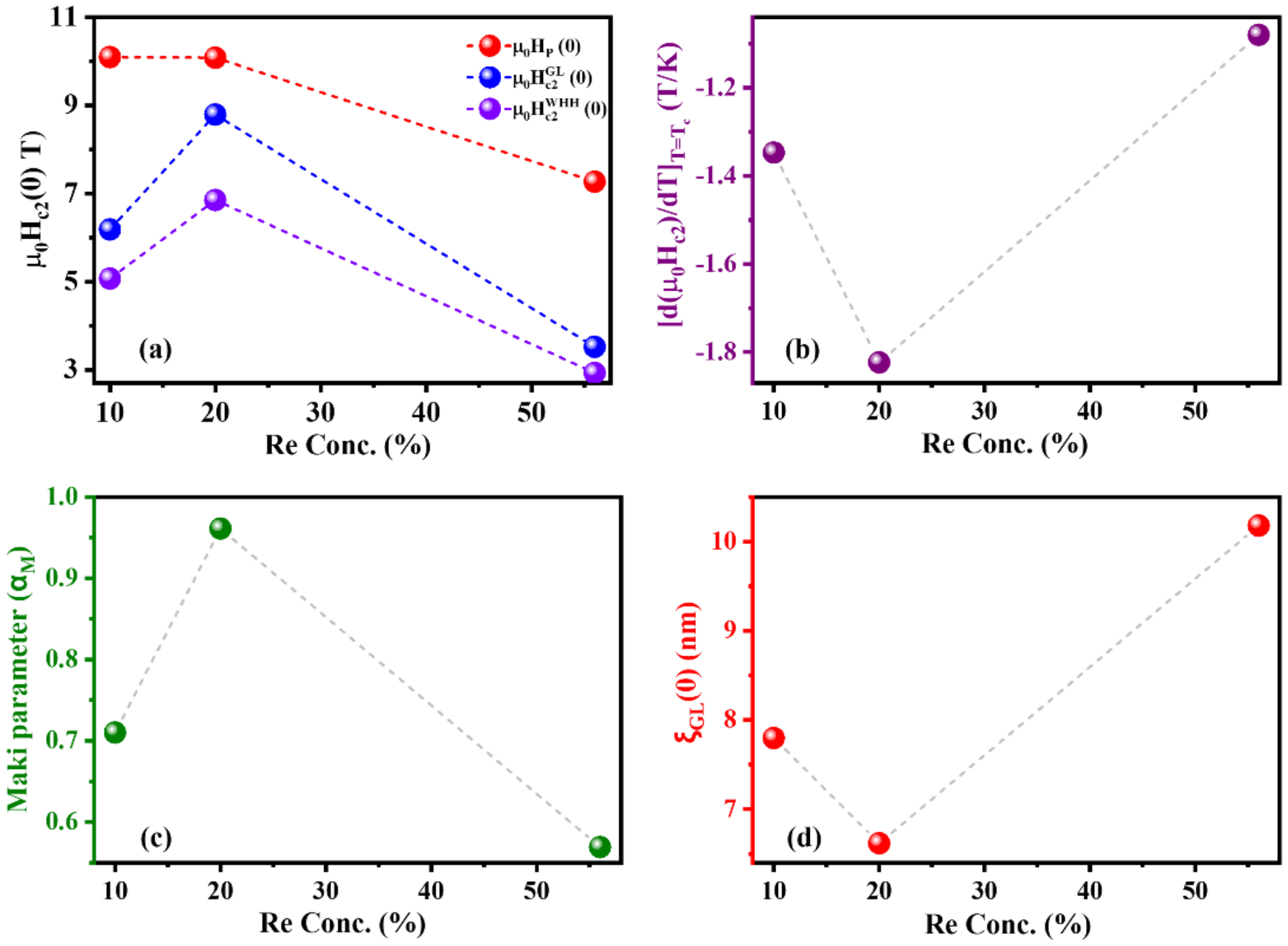

***Figure 3:*** ***(a)*** *shows a comparison plot of* $\mu_0 H_{c2}^{P}(0)$ , $\mu_0 H_{c2}^{GL}(0)$ & $\mu_0 H_{c2}^{WHH}(0)$ *for all three samples,* ***(b)*** *Slope value of* $d(\mu_0 H_{c2})/dT$ *at* $T=T_c$*,* ***(c)*** *Maki parameter (*$\alpha_M$*) calculated and plotted against Re. concentration %,* ***(d)*** *Coherence length vs Re. concentration %.*

To assess the possible pair-breaking mechanism, the Maki parameter ($\alpha_M$) has been calculated using the relation, $\alpha_M = \sqrt{2}\mu_0 H_{c2}^{WHH}(0)/\mu_0 H_{c2}^{P}(0)$ [43]. The values are plotted against their concentrations in Fig.3(c), where RNZHT 10 & 56 are well below 1, implying the dominance of orbital pair breaking over the Pauli paramagnetic effect. However, in RNZHT 20, $\alpha_M$ is close to the value 1, the upper critical field $\mu_0 H_{c2}^{GL}(0)$ approaching the Pauli limit $\mu_0 H_{c2}^{P}(0)$ suggesting that, while orbital pair breaking remains the primary mechanism limiting superconductivity, Pauli pair breaking may become non-negligible as a secondary contribution. Further, the Ginzburg-Landau coherence length ($\xi_{GL}(0)$) has been calculated using the relation $\mu_0 H_{c2}^{GL}(0) = \Phi_0/2\pi\xi_{GL}^2$, where $\Phi_0$ is the flux quantum (2.067833848 x $10^{-15}$ Wb), and these values are plotted against the Re concentration in Fig.3(d). RNZHT 20 exhibits the shortest coherence length, consistent with its largest upper critical field among the studied compositions.

## (c) Magnetization measurements:

Temperature-dependent dc magnetization has been measured in zero-field-cooled warming (ZFC) and field-cooled cooling (FC) modes under a magnetic field ($\mu_0 H$) of 2 mT near the $T_c$ region of the samples, and the results are depicted in Fig.4(a-c). A strong diamagnetic signal indicating the onset of $T_c$ is observed for all three compositions and plotted in Fig.4(d), and it is consistent with the $T_c$ obtained from resistivity. Additionally, the separation between the ZFC and FC curves demonstrates magnetic irreversibility consistent with flux pinning, with larger separation for RNZHT 10 and 20 than for RNZHT 56. RNZHT 56 also has a relatively small $\mu_0 H_{c2}(0)$. Further, the pronounced separation below 2.5 K and the magnetization hysteresis provide mutually consistent evidence for enhanced low-temperature vortex pinning in RNZHT 20.

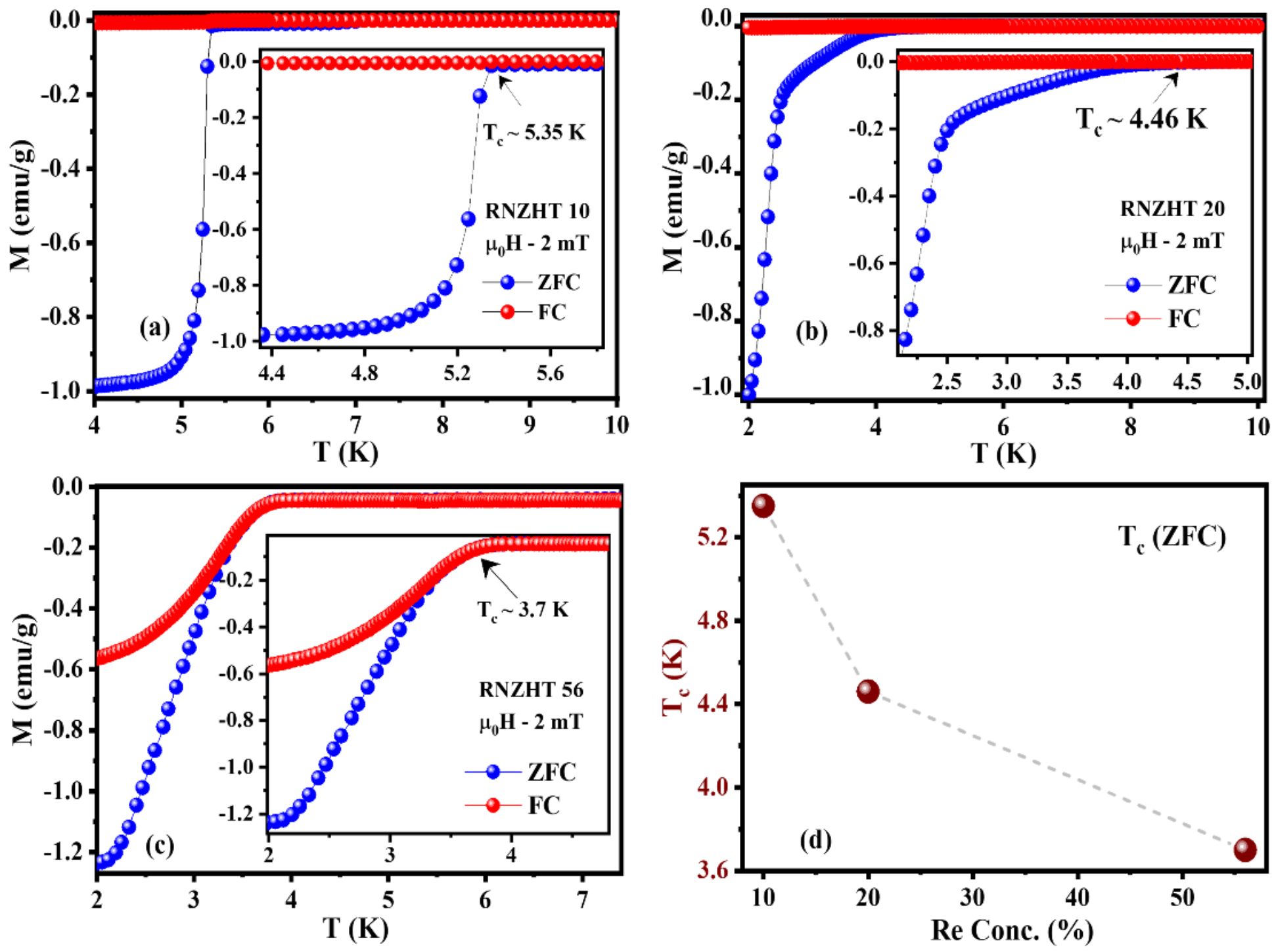

***Figure 4: (a) – (c)** show the temperature-dependent DC magnetization at a fixed external field ($\mu_0 H$) of 2 mT in both ZFC and FC modes near the $T_c$ region; **(d)** shows the transition temperature $T_c$ vs Re. concentration %.*

In order to understand more about the vortex state, the field-dependent isothermal magnetization, $M(\mu_0 H)$, at various fixed temperatures for the samples RNZHT 10 & 20 was measured and is shown in Fig.5(a) & 5(b). The lower critical field $\mu_0 H_{c1}$ at different temperatures was determined from the point of deviation of the Meissner signal from its linearity for each measured temperature for RNZHT 10 and 20, and it is shown in the insets of the $M(\mu_0 H)$ figures. For RNZHT 20, as shown in the inset of Fig.5(b), the Meissner signals at 2 K are stronger than above 2 K, which can be correlated with the flux pinning observed at low temperatures, similar to the report on Hf-Nb-Ta-Ti-V [44]. This behavior is consistent with stronger low-temperature magnetic irreversibility in RNZHT 20. The estimation of $\mu_0 H_{c1}$ at 0 K, $\mu_0 H_{c1}(0)$, has been calculated using the GL equation by fitting the temperature-dependent $\mu_0 H_{c1}$ values with $\mu_0 H_{c1}(T) = \mu_0 H_{c1}(0)\,[\,1 - (\frac{T}{T_c})^2\,]$ [45]. The fitted data are depicted in Fig.5(c-d), and the calculated $\mu_0 H_{c1}(0)$ values are plotted against Re concentration and shown in Fig.5(e). For RNZHT 56, it was taken from the previously reported values [31].

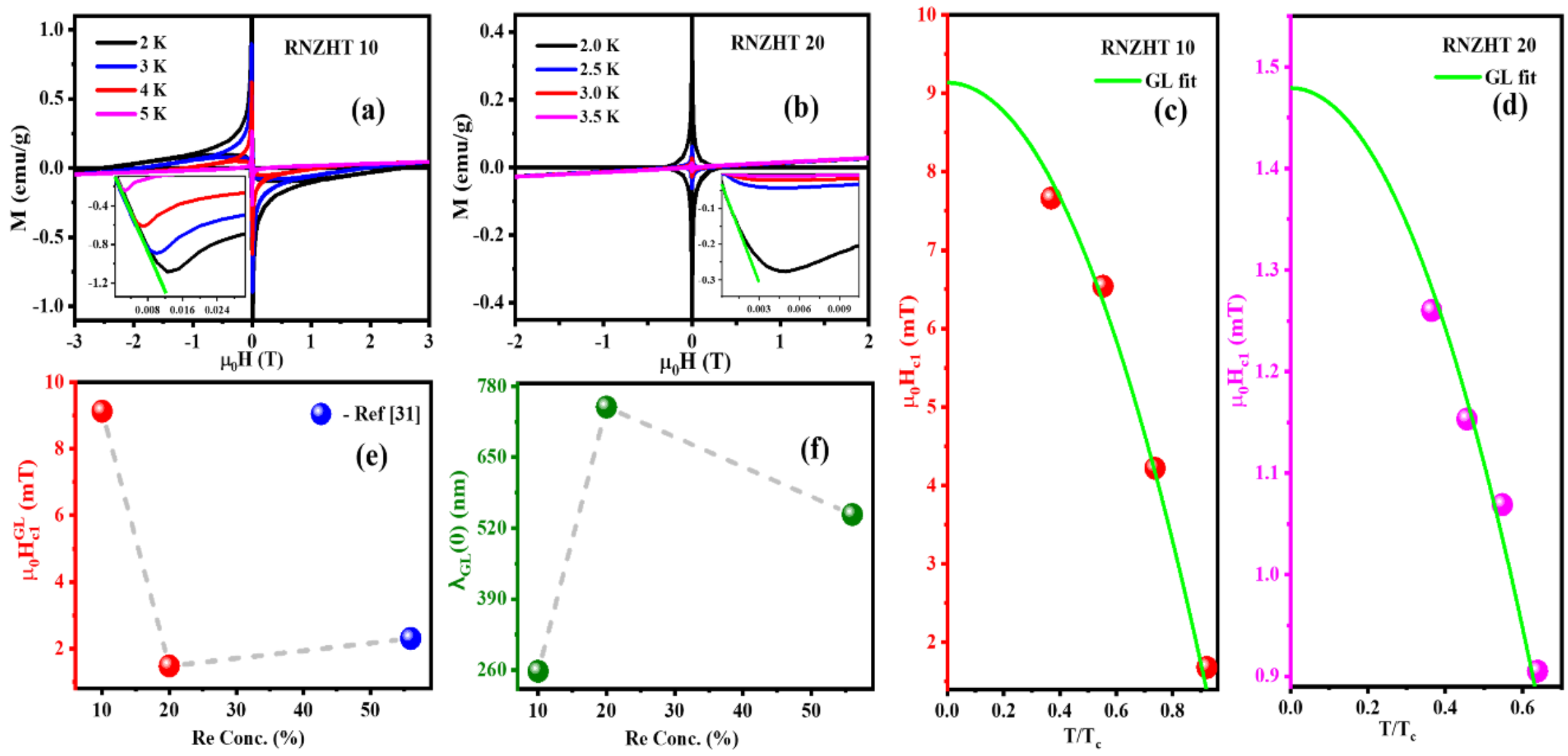

***Figure 5:*** ***(a-b)*** *shows the field ($\mu_0 H$) dependent magnetization (M) at various fixed temperatures within $T_c$ region with insets showing the magnified Meissner signal,* ***(c-d)*** *depicts the GL relation fitted on $\mu_0 H_{c1}(T)$ vs $T/T_c$,* ***(e-f)*** *shows the lower critical field* $\mu_0 H_{c1}(0)$ (Here, $\mu_0 H_{c1}(0)$ for RNZHT 56 is adapted from [31]), *Penetration depth ($\lambda_{GL}(0)$) vs Re. concentration, respectively.*

In addition to the coherence length, the GL penetration depth ($\lambda_{GL}(0)$), was estimated using the equation $\mu_0 H_{c1}(0) = \frac{\Phi_0}{4\pi\,\lambda_{GL}^2(0)}\left[\ln\left(\frac{\lambda_{GL}(0)}{\xi_{GL}(0)}\right) + 0.12\right]$. The obtained values are plotted in Fig.5(f), with RNZHT 20 exhibiting the largest penetration depth among the investigated samples. The upper critical field is governed primarily by the coherence length rather than by the penetration depth. To further characterize the superconducting state, the GL parameter was calculated using $\kappa_{GL} = \frac{\lambda_{GL}(0)}{\xi_{GL}(0)}$ with the extracted length scale parameters ($\xi_0$ & $\lambda_{GL}$). A type II superconductor satisfies $\kappa > 1/\sqrt{2}$ [46,47]. The resulting values are 35.31, 121.26, and 56.37 for RNZHT 10, 20 & 56, respectively. These values are considerably larger than the threshold, confirming that all three samples are strongly type II superconductors. Finally, the thermodynamic critical field $\mu_0 H_c$ (0) was calculated using the GL relation $\mu_0 H_{c2}(0) = \sqrt{2}\kappa_{GL}(\mu_0 H_c(0))$, yielding values of 123.75, 51.25 and 44.15 mT for RNZHT 10, 20 & 56, respectively.

**(d) Cocktail effect and VEC:**

To explain the cocktail effect in our samples, the compositionally averaged counterpart of $T_c$ is calculated using the relation, $\bar{X} = \sum_{i=1}^{n} c_i X_i$, where X is the parameter. In our case, it is the superconducting transition temperature ($T_c$). The calculated $\overline{T_c}$ is shown in the given Table 1. The ratio between the $T_c$ and the $\overline{T_c}$ increases monotonically. This ratio indicates that the measured $T_c$ of the studied HEA exceeds the composition-weighted average of the elemental $T_c$. Although the ratios are not as high as the reported value for other Re-based HEAs [48], the measured $T_c$ still shows a slight enhancement. A ratio greater than unity indicates a positive deviation from a simple composition-weighted average of elemental transition temperatures [49]. By this conventional composition-weighted criterion, all three alloys exhibit a positive cocktail effect that strengthens with increasing Re concentration and reaches its largest value in hcp RNZHT 56. Further, the valence electron count (VEC) or e/a is a crucial parameter in stabilizing the crystalline solid solutions to their favorable crystallographic phases. It is calculated for all three samples using the following relation [48], $VEC = \sum_{i=1}^{n} c_i (VEC)_i$, where $(VEC)_i$ is the valence electron count for compositional element i, and $c_i$ is the atomic fraction of element i. The calculated values are presented in Table 1, where VEC increases with the concentration of Re in the composition, which is similar to the report [50]. The systematic decrease in $T_c$ with increasing VEC, accompanied by the evolution from bcc toward hcp structure, strongly supports a coupled electronic, lattice, and structural origin rather than a simple atomic-mass effect. Figure 6 shows the critical temperature of the samples plotted as a function of VEC.

For comparison, the values are plotted against the critical temperature for 4d transition metals and alloys in their crystalline form (green region shaded under the green line) as well as their amorphous vapor-deposited film form (red region below the red line). This green line, commonly referred to as the Matthias rule, shows maxima $T_c$ around the VEC values of ~4.7 and ~6.8 with a non-monotonic variation [35]. The red dotted line serves as the boundary for superconductivity with a linear increment trend up to the maximum VEC of ~6.6 for 4d amorphous alloys explored by Collver and coworkers [36,51]. These two trend lines act as the benchmark of comparison for other superconductors. Along with these, other reports on bcc [20,21,52–56] and hcp [57–59] structured HEA superconductors are also presented. Typically, most of the bcc samples crystallized under the 4.2 ≤ VEC ≤ 5.2 regime and hcp systems in the range of 7.1 ≤ VEC ≤ 7.8. In RNZHT 10 and 20 phase stabilization, the VEC value falls under the reported range of bcc, and also a decrement in $T_c$ with an increase in VEC, similar to crystalline 4d alloys, can be observed. But the results get interesting while comparing the hcp sample, RNZHT 56, having VEC closer to the bcc reported range rather than the hcp range. This deviation could be attributed to a robust 5d orbital hybridization in the sample with a higher Re fraction.

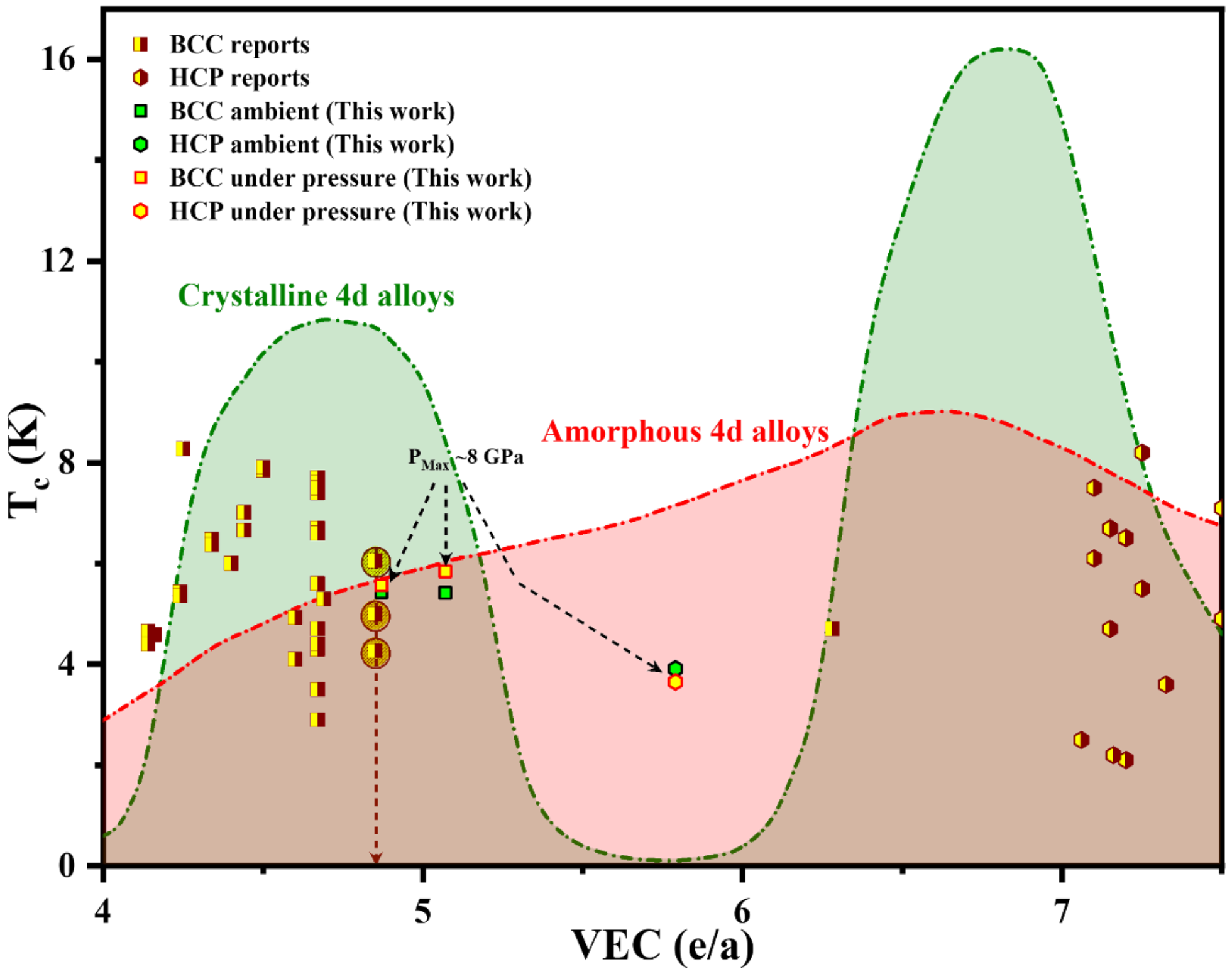

***Figure 6***: *showing $T_c$ vs VEC for previously reported bcc and hcp structures in comparison with the values of the present work, including pressure results.*

**Table 1:** HEA parameters (VEC, $T_c$, $\overline{T_c}$ , and $\frac{T_c}{\overline{T_c}}$) for the sample RNZHT 10, 20, & 56 are given below.

| **Concentration of Re** | **Phase** [31] | **VEC** | $T_c$ **(K)** | $\overline{T_c}$ **(K)** | $\frac{T_c}{\overline{T_c}}$ |
|---|---|---|---|---|---|
| **10** | bcc | 4.87 | 5.423 | 5.30 | 1.02 |
| **20** | bcc | 5.10 | 5.419 | 4.60 | 1.18 |
| **56** | hcp | 5.79 | 3.907 | 2.04 | 1.93 |

## 2. Transport properties under physical pressure:

Figure 7(a-c) shows the *ρ(T)* under hydrostatic pressure conditions up to ~8 GPa for RNZHT 10, 20 & 56 samples. A sluggish decrease in resistivity with temperature indicates a poor metallic nature of the samples. Across the entire pressure range, each sample maintained its metallicity and superconductivity. With increasing pressure, only a marginal decrease in the normal-state resistivity ($\rho_{NS}$) is observed, which is consistent with other reports on HEAs [26,28,60] indicating only modest changes in normal-state transport. The normal-state resistivity may also depend on the Debye temperature ($\Theta_D$) and density of states (DOS), as reported for the quaternary TiZrHfNb HEA under compression [61]. Although a monotonic reduction in normal state resistivity is observed for every studied sample, $T_c$ increases monotonically for RNZHT 10 and 20, whereas RNZHT 56 shows an opposite trend with applied external pressure, as shown in Fig.7(e). In general, the superconducting transition temperature results from the competition between the logarithmically averaged characteristic phonon frequency ($\omega_{log}$) and the electron–phonon coupling constant ($\lambda$). The predominance of one factor over the other leads to either an enhancement or suppression of $T_c$ with applied pressure [28], which may be consistent with the observed trend.

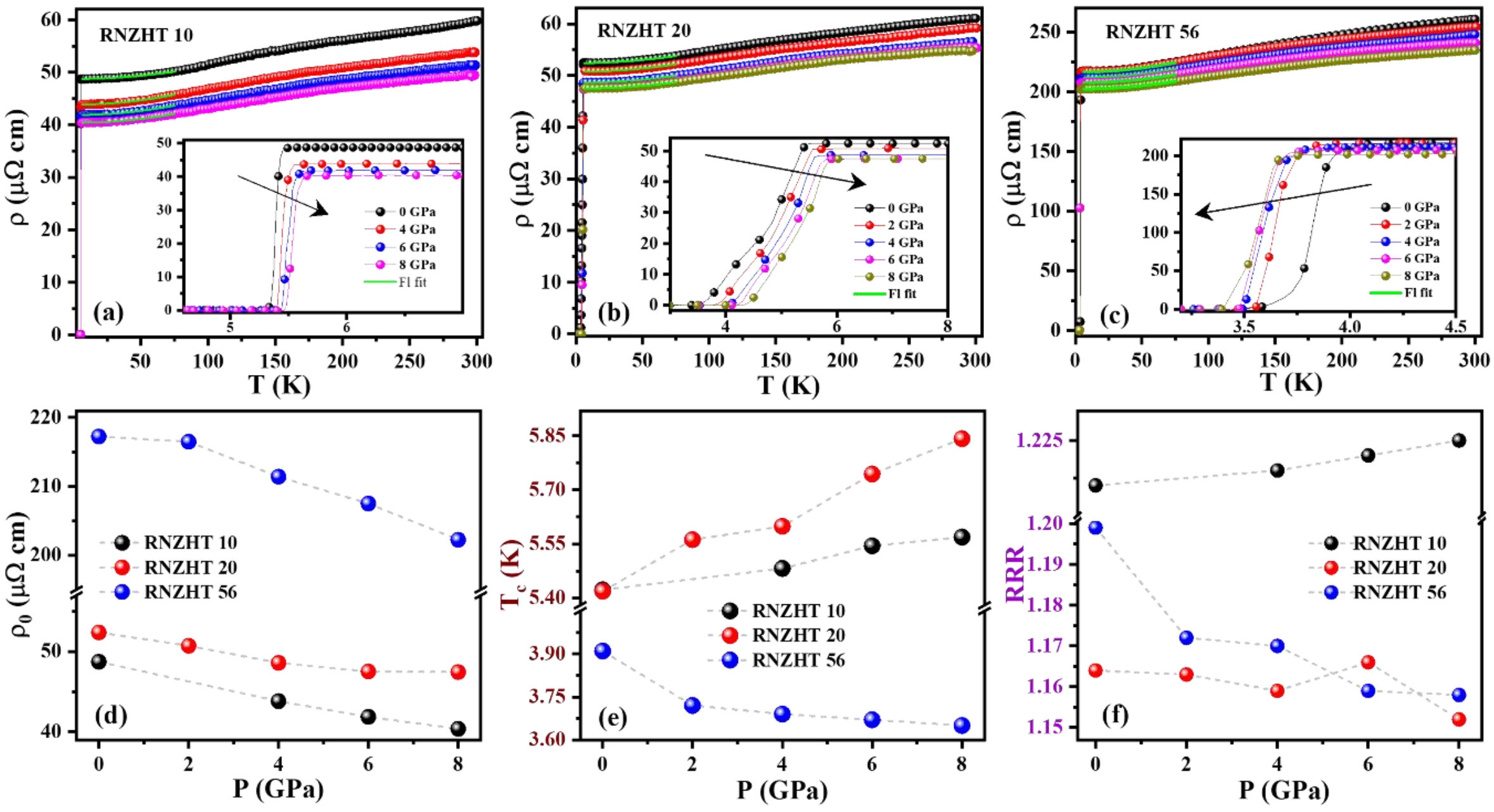

***Figure 7: (a-c)*** *show the zero-field temperature dependent resistivity (ρ(T)) at various fixed hydrostatic pressures together with Fermi-liquid fit (bright green solid line) and insets showing the transition region;* ***(d)*** *represents the evolution of residual resistivity* $\rho_0$ *with pressure,* ***(e)***

*shows the pressure dependence of $T_c$ and the corresponding linear fits for all samples, **(f)** shows the RRR values for all the samples under pressure.*

The reversal in the sign of $dT_c/dP$ between the bcc and hcp members is the central pressure result of this study and establishes a robust structure-associated superconducting response within a single alloy series. As mentioned in the introduction, RNZHT 10 & 20 follow the bcc structure, and RNZHT 56 follows the hcp structure. Although pressure can substantially modify the electronic structure and produce dome-shaped $T_c(P)$ behavior in bcc HEA superconductors, their crystal structure can remain stable to very high pressure (~190 GPa) [26,28]. A recent study of $Re_{0.60}(NbTiZrHf)_{0.40}$ reported an hcp-to-bcc transition near 44 GPa, approximately five times the maximum pressure of ~8 GPa applied to RNZHT 56 in the present work. That transition was interpreted as a diffusionless shear process energetically stabilized by the pressure-induced rearrangement of the electronic states. Figure 6 shows that most of the bcc HEA superconductors possess higher $T_c$ with VEC values comparatively closer to the VEC of our hcp structured $Re_{0.56}(NbZrHfTi)_{0.44}$. The hcp-to-bcc transition reported near 44 GPa for the closely related $Re_{0.60}(NbTiZrHf)_{0.40}$ [29] provides a compelling motivation for extending the present measurements beyond 8 GPa, where a qualitative change in the superconducting response of RNZHT 56 may emerge.

In addition, linear fits to the pressure dependence yield $dT_c/dP$ values of approximately 0.018 K/GPa, 0.053 K/GPa & -0.033 K/GPa for RNZHT 10, 20, & 56, respectively, with RNZHT 20 exhibiting the largest positive slope. Similar to the ambient condition, the Fermi liquid relation is used to fit the baric evolution of $\rho(T)$, and the extracted $\rho_0$ values are plotted in Fig.7(d). All three samples show a monotonic decrease in $\rho_0$, with RNZHT 20 showing the smallest magnitude of $d\rho_0/dP$ and a tendency toward saturation at higher pressure. The pressure-dependent RRR factor for the temperature range of 295 to 10 K has been calculated and shown in Fig. 7(f), showing that RRR changes only modestly over the investigated pressure range. Further, Fig.6 highlights that the pressure dependence of $T_c$ for all three samples is similar to that of the isoelectronic samples highlighted near the VEC value of ~4.85, in which enhanced electron-phonon coupling, optimized DOS at the Fermi level, and phonon softening are listed among the $T_c$ enhancing mechanisms, similar to our earlier argument. Similar mechanisms have been discussed for chemically tuned HEAs [21]. Within a phonon-mediated pairing picture, the opposite pressure slopes strongly suggest structure-dependent changes in the balance among the density of states, characteristic phonon frequencies, and electron-phonon coupling. The fitted residual resistivity decreases with pressure for all three samples, suggesting reduced low-temperature resistive scattering, although pressure-induced changes in sample geometry and electronic structure cannot be excluded.

**Summary & conclusion:**

In summary, $[Nb_{0.67-x}Re_x][TiZrHf]_{0.33}$ provides a single chemically related HEA series in which composition, crystal structure, disorder, and physical pressure can be compared directly. According to the conventional composition-weighted criterion, all three compositions exhibit a positive cocktail effect, which increases with Re concentration. RNZHT 20 is the most distinctive composition: its broad transition, minimum RRR, largest upper critical field and GL parameter, Maki parameter close to unity, and largest positive $dT_c/dP$ consistently identify it as the composition most strongly affected by disorder and proximity to the reported structural crossover. The central finding is a structure-selective pressure response of superconductivity. $T_c$ increases under pressure in the bcc RNZHT 10 and 20 samples but decreases in the hcp RNZHT 56 sample. This reversal in the sign of $dT_c/dP$ establishes a robust structure-associated contrast and identifies crystallographic structure as a key organizing variable governing the superconducting response. Metallic transport and superconductivity remain robust up to approximately 8 GPa in all three samples, while the opposite pressure slopes indicate that chemical substitution and physical compression are complementary but nonequivalent tuning routes. The reported hcp-to-bcc transition near 44 GPa in the closely related $Re_{0.60}(NbTiZrHf)_{0.40}$ alloy motivates higher-pressure measurements of RNZHT 56, where a qualitative change in its superconducting response may emerge.

**Acknowledgments:**

T.M. and S.A. acknowledge the Board of Research in Nuclear Sciences (BRNS) for supporting the research through the research Grant 58/14/06/2021 - BRNS/37097. Y.U. acknowledges JSPS KAKENHI Grant Number JP25K00951. M.N.S gratefully acknowledges the receipt of a fellowship from the ICTP Programme for Training and Research in Italian Laboratories (TRIL), Trieste, Italy. R. P. S. acknowledges the SERB, Government of India, for the Core Research Grant No. CRG/2023/000817. S.M. acknowledges ISSP, The University of Tokyo, Japan, for providing the experimental facilities during the visit.